# THE TRANSFORMATION RISK-BENEFIT MODEL OF ARTIFICIAL INTELLIGENCE: BALANCING RISKS AND BENEFITS THROUGH PRACTICAL SOLUTIONS AND USE CASES

Richard Fulton,[1] Diane Fulton,[2] Nate Hayes[3] and Susan Kaplan[3]

[1] Department of Computer Science, Troy University, Troy, Alabama, USA
[2] Department of Management, Clayton State University, Morrow, Georgia, USA
[3] Modal Technology, Minneapolis, Minnesota, USA

## ABSTRACT

*This paper summarizes the most cogent advantages and risks associated with Artificial Intelligence from an in-depth review of the literature. Then the authors synthesize the salient risk-related models currently being used in AI, technology and business-related scenarios. Next, in view of an updated context of AI along with theories and models reviewed and expanded constructs, the writers propose a new framework called "The Transformation Risk-Benefit Model of Artificial Intelligence" to address the increasing fears and levels of AI risk. Using the model characteristics, the article emphasizes practical and innovative solutions where benefits outweigh risks and three use cases in healthcare, climate change/environment and cyber security to illustrate unique interplay of principles, dimensions and processes of this powerful AI transformational model.*

## 1. INTRODUCTION

While Artificial Intelligence (AI) offers many benefits in improving efficiency, accuracy, accessibility, pattern-recognition and creating higher-paid highly skilled technology-related jobs, sustainability and quality of life, it also presents risks including invading privacy, displacing jobs, creating bias, increasing fraud/deception and weaponizing of AI. As the costs and number of risks associated with AI rise, the ability both to assess those risks and to engage workers in implementing controls will be the new competitive advantage [1].

This article addresses the risks of AI through the lens of benefit-risk analysis, based on relevant models and theories. From an analysis of AI advantages and risks, the authors review risk models to create a new paradigm for transforming AI risks into benefits. A set of principles, dimensions and processes that work in the AI context are laid out from thorough analysis of previous work, addressing the limitations and omissions of empirical risk models. Next, innovative and practical solutions are based on principles, dimensions and processes of risks and opportunities. Lastly, three use cases in healthcare, climate change/environment and cyber security add depth to the body of knowledge and help leaders visualize how to implement balancing risks with benefits.

## 1.1. Benefits of Artificial Intelligence

From the industrial revolution, significant development in technical innovation has succeeded in transforming manual tasks and processes that had been in existence for decades where humans had reached the limits of physical capacity. It is no surprise that AI offers this same transformative potential for the augmentation of human tasks within a wide range of industrial, intellectual and social applications [2]. If society and organizations managed the risks associated with the industrial revolution technologies, it is apparent that addressing the challenges of AI is also within reach.

Faced-paced breakthroughs in algorithmic machine learning and autonomous decision-making give leaders new opportunities to balance risks and benefits[2]. Through data analysis, identifying patterns and making predictions, AI can help society to solve difficult problems like cybercrimes and create a better world, impacting education, healthcare, clean water, climate change, poverty and hunger. By harnessing the power of AI, society can create a worldwide sustainable future [3]. In fact, from a comprehensive analysis study, AI was found to be an enabler on 134 general targets (79%) across the areas of Sustainable Development Goals (SDGs) set by the United Nations aimed at overcoming certain pressing worldwide problems.

In the societal and environmental areas of challenges, AI can enhance the provision of food, health, water and energy services to the population and create smart and low-carbon cities encompassing a range of interconnected technologies such as electric autonomous vehicles and smart appliances that enable demand response in the electricity sector and smart grids that match electrical demand to weather conditions [4]. By 2030, experts predict 20 to 50 million new AI-related jobs will be created globally and overall spending on technology could increase by more than 50 percent by 2030 [5]. Studies have analyzed the societal, economic and organizational impact of this significant change, showing a changing jobs market that is predicted to focus humans on more creative and higher thinking skills in AI-support roles that pay more [6]. Generative AI is creating more types of jobs [7]. Upwork Research Institute's new survey of 1,400 U.S. business leaders confirms the positive job impact of AI. Although there may be variation in generative AI adoption of technologies such as ChatGPT and Midjourney across companies, two-thirds of top leaders agree that they will increase hiring due to generative AI [8].

The benefits of applying AI technologies to big data problems are their analytic insight and predictive capability [9]. Health related studies, for example, show AI technologies can greatly support patient health-based diagnosis and predictive capability [10][11][12][13]. Economically, AI generates productivity gains [14]. In the legal and policy arena, the best cybersecurity solutions have AI and machine learning built into them, which creates dynamic protection to face ever-improving attacks and fraud threats that use automated scripts[15]. Although generative AI helps those who wish to radicalize people to create deep fake videos used online in social media, AI tools can also prevent it from succeeding. In November 2022, Intel released FakeCatcher, a cloud-based AI tool that it claims can accurately detect fake videos 96% of the time, using up to 72 different detection streams [16].

Retail and service-oriented organizations benefit from AI's ability to personalize the online experience through real-time, tailored product recommendations based on customers' purchase history, age, gender, geographic location, and other data points which continuously upgrades its capabilities by analyzing millions of customer-service interactions in real time [17]. With the help of AI, companies can save time and resources and can create even more personalized experiences directly communicating with their customers, which will result in enhanced brand loyalty and lifetime relationships [17]. Item recommendation is instrumental for a content provider to grow its audience and evidence shows that automated recommendations account for 35% of sales on amazon.com, 50% of initial messages sent on match.com and 80% of streamed hours on netflix.com[18]. Big Data Analytics (BDA) develops the methodological analysis of large data structures tackling data volume, velocity, variety, veracity and value issues. BDA combined with AI has the potential to transform areas of manufacturing, health and business intelligence offering advanced insights within a predictive context [19][20][21]. Organizations are increasingly deploying data visualization tools and methods to make sense of their big data structures. Because the analysis and processing of complex heterogeneous data is problematic and human perception and cognition maybe limited, organizations can extract significant value, understanding and vital management information from big data via intelligent AI-based visualization tools [22][13][23]. AI benefits are organized into societal (SOC), economic (ECON), ethical (ETH), political/ legal/policy (POL), environmental (ENV), data (DATA), technological/implementation (TECH) and organizational/managerial (ORG) arenas [13] and examples in each category are shown in Table 1.

## 2. RISKS OF ARTIFICIAL INTELLIGENCE

The implementation of AI technologies can present significant challenges or risks for governments and organizations as the scope and depth of potential applications increase and the use of AI becomes more mainstream. Although AI-based systems are increasingly being leveraged to provide value to organizations, individuals and society, significant attendant risks have been identified recently. [24] divided risks into two categories including 1) *Risks to fundamental rights*—flaws in the overall design of AI systems or biased data can lead to breaches in fundamental rights, including free speech, discrimination based on sex, race, religion, disability, age or sexual orientation, protection of personal data and private life and consumer protection; and 2) *Risks to safety and liability*—flaws in the design of AI technologies may present new safety risks for users related to the availability and quality of data or other problems stemming from AI and machine learning.

This research builds upon 8 AI challenge categories proposed in a public sector model, used to organize the risks: 1) societal; 2) economic; 3) political, legal and policy; 4) environmental; 4) ethical; 5) data; 6) technological and implementation; and 8) organizational and managerial [13]. Public organization adoption of AI and data science presents numerous known challenges ranging from employee path dependency on embedded processes and norms, information silos, and a lack of resources, collaborative culture and technical capacities [25][26][27]. AI systems have had some harmful side effects on communities, through effects on employment and inequality [28], privacy and safety—injury, property loss and workplace hazards [14], addictive behavior [29], fairness, bias and discrimination [30][31], human rights [12] and polarization, extremism, manipulative practices and conflict [32][33][34][18].

For example, researchers found race bias [35][36] and gender bias [37] for policing algorithms and bias against working-class and disadvantaged communities for educational assessments in the UK during COVID [38]. In addition, an AI algorithm created by Zillow was unable to accurately capture complex assessments such as estimating home values thus causing layoffs of 2,000 employees and a sell-off of its iBuying division [39]. Industry-friendly hackers also fooled Tesla's Autopilot AI program into merging into oncoming traffic and took control of the car using a video game controller [40]. To make matters worse, the high legal bar to prove either a disparate treatment or disparate impact cause of action under Title VII, coupled with the "black box" nature of many automated hiring systems makes the detection and redress of bias exceedingly difficult [30]. In the intersection of privacy and human rights, AI used for target marketing and customer service may gather data that may include the user's private behaviors as playing a certain game, smoking, watching porn or defaulting on loans [18]. Concerns regarding online personalization range from recommendation algorithms isolating information seekers from differing viewpoints (filter bubble), radicalizing citizens' attitude towards controversial issues (polarization) or enabling malicious content [18]. Generative AI has become widely popular, but its adoption by businesses comes with a degree of ethical risk [41]. With relatively modest amounts of data and computing power using generative AI, the average person can create a video of a world leader confessing to illegal activity leading to a constitutional crisis, a military leader saying something racially insensitive leading to civil unrest in an area of military activity, or a corporate titan claiming that their profits are weak leading to global stock manipulation. These so called 'deep fakes' pose a significant threat to our democracy, national security and society [32]. There are also practical issues over how accurate machine learning solutions are. The range of testing approaches available within machine learning is growing rapidly, and that is a good thing, but it is also driven by the evident limitations of the previous methods and the need to overcome those limitations [42].

Table 1. Salient AI benefits by categories and examples.

| Categories | Examples |
|---|---|
| SOC | Worldwide productivity gains; Workers can choose what tasks they want to do; Societal increased well-being; Impact world sustainability goals of UN; Analytics and AI based predictions create better patient healthcare; Improve Customer/Organization Interface; Tackle world problems; Overcome Cultural barriers; Protect Human rights |
| ECON | Increase higher-skill and higher-paying jobs; Create new job categories; Improve profitability; Optimize returns on investment |
| POL | Detect deep fakes, propaganda, or spying on users; Fight cybercrimes and fraud; Stop weaponizing AI for harmful purposes; Add rules and accountability in the use of AI; Rely on European Union guidelines/rules for responsible AI |
| ENV | Address climate change and resources scarcity; impact carbon footprint. |
| ETH | Build trust towards AI-based decision making; Increase explanation of decisions made by AI; improve processes relating to AI and human behavior, find balance in machine versus human value judgements; be vigilant in testing for bias/discrimination |
| DATA | Advanced insights within a predictive context; improve 5 V's of data; Increase transparency and reproducibility; Create reliable/sufficient data pools; Improve data integration and continuity; Create standards for data collection |
| TECH | Fight adversarial attacks; Increase transparency and interpretability; Smart design of AI systems; Increase AI safety and security; gain value from big data; continue to innovate ways to decrease problems using unstructured data; integrate legacy and new systems through collaboration with vendors, governmental and educational institutions |
| ORG | Personalize customer experiences; Streamline repetitive and boring jobs; Visualize to increase understanding of firm needs; Improve data sharing and collaboration; Increase AI talent through HR innovation in hiring and retraining; Create an AI-driven culture |

Another crucial factor is the availability of suitable data. Although machine learning packages for Python and R can easily read all types of data from Excel to SQL and can perform natural language processing and process images, the speed with which machine learning solutions have been proposed has not kept pace with firms' abilities to suitably organize the internal data they have access to. Data is often held in separate silos across departments, on different systems and with internal political and regulatory issues restricting the sharing of data. Important data might not even be recorded as data but rather kept as informal knowledge of the firm [42]. Estimates for work displacement due to automation highlight that up to a third of current work activities (between 400 million and 800 million jobs around the world) could be impacted by 2030 [5]. A survey of the top 1000 firms in the United States found that their biggest concern in the implementation of AI was the readiness and ability of staff to understand and work with these novel solutions [43]. Sixty-four percent of US executives and 70 percent of European leaders believe they will need to retrain, up-skill or replace a fourth of their workers due to advancing automation and digitization [44]. Please see Table 2 for a summary of salient AI risk examples by categories.

## 3. MODELS AND THEORY

Very few comprehensive models of AI risk are available today to help managers assess and deal with the increasing risks they face. To address this void and the ever-increasing depth and breadth of AI technology challenges, the researchers critically review existing related risk models and start with an historical empirically-based risk model dealing with the adoption of information technology projects. Since the risk of IT projects failing was extremely high in those days, [45] developed a model to predict project risk and delineate risk factors associated with organizational IT projects, which serves as a basis for a newly proposed AI risk framework.

The McFarlan Risk Model (MRM) provided a useful and measurable approach for the diagnosis and mitigation of IT project risks with three dimensions based on 'project size', 'project structure' and 'experience with technology'. For example, if there were excessive costs, large numbers and levels of staff needed, increased completion time to complete and impacted many different departments, then the risk was increased. If the user department needed to change a lot of procedures and structures to meet the project requirements and users were highly resistant to changes, the project was also considered higher risk. Lastly, if the team lacked the appropriate experience with the recent technologies, the firm needed to hire more experts or use outside consultants and if the experts did not work in partnership with the company, the risks expanded exponentially [45].

These three dimensions were expanded in this research to include 'resources' (not just the size of a project but how many, how well and how long resources support AI implementation), 'governance' including guidelines and structural processes put in place (not just the project structure) and 'expertise building' or on-going capacity and commitment to train, retrain, upskill and encourage employee development (not just relying on the current experience of the workers or hiring outsiders). In their "Risk Assessment Framework" on implementing enterprise resource planning projects, risks from 'external engagement', program management, work stream and work package levels across technical, operational, business and organizational categories were successfully mitigated using on-going risk controls [46].

The specific content of corporate governance guidelines and policies is an important variable to any risk mitigating model. Using a comparison between Belgium (weaker country guidelines) and Australia (stronger governance guidelines), researchers found significantly more developed risk management and internal control systems in Australian companies versus Belgian companies overall [47]. In addition to strong governance guidelines, impactful 'governance' also dictates that an entire risk management structure be in place. For example, the structure of risk control may include 4 phases of identifying, assessing, mitigating and monitoring risks [48] or 4 specific functions— govern, map, measure and manage — to help organizations address the risks of AI systems [49].

In this research, the 'governance' dimension of risk focuses on a broader range of AI policies, controls and structure. Using a comprehensive AI risk management system can maximize the benefits of AI technologies while reducing the likelihood of negative impacts to individuals, groups, communities, organizations and society" [49]. In their survey of 576 companies and a review of 2,750 company/analyst reports, researchers found that financial performance was highly correlated with the level of integration and coordination across risk, control and compliance functions. Effectively harnessing AI technology to support risk-benefit management is the greatest weakness or opportunity for most organizations [50]. To improve the predictive ability of McFarlan's model, the "Extended McFarlan Risk Model" (EMRM) was developed to differentiate between project success and failure, adding an organizational 'culture' dimension [51]. The 'culture' attributes were quantitatively measured including users' practices, users' attitudes, company working practices, organizational polices such as information technology policy and data flow practices, internal and external communication practices in the organization, openness to change and cross-functional coordination. If the corporate culture was lacking, the implementation risk increased [51]. In this research, the dimension of 'culture' was widened to encompass an 'AI-driven culture' with communication, work policies/practices, change management, innovation and cross- disciplines/perspectives which fully integrate AI throughout the firm for triple bottom line goals of profits, people and the planet [52]. The new risk-benefit model proposed is also based on the use of social cognitive theory which views people as active agents who can both influence and can be influenced by their environment [53].

Table 2. Salient AI risks by categories and examples.

| Categories | Examples |
|---|---|
| SOC | Customer/Organization Interface; World crises; Cultural barriers; Human rights; Country specific data profiles; Unrealistic expectations towards AI technology; Country specific organizational practices and insufficient knowledge on values/advantages of AI. |
| ECON | Affordability of required computational expenses; High costs for customers; High cost and reduced profits for organizations; Wider divides in society leading to social upheaval |
| POL | Copyright issues; Embedded bias and discrimination by humans or technology; Injury; Governance of autonomous intelligence systems; Responsibility and accountability; reduced privacy/safety; National security threats from foreign-owned companies/governments; Cybercrimes; Weaponizing AI for harmful purposes; Lack of rules and loss of accountability; Costly human resources still legally required to account for AI based decision; Lack of official industry standards of AI use and performance evaluation; difficult to redress bias legally |
| ENV | Impact on carbon footprint, use of resources such as fossil fuels, water, electricity; impact on climate change and air quality |
| ETH | Lack of trust towards AI based decision making and unethical use of shared data. Responsibility and explanation of decision made by AI; processes relating to AI and human behavior, compatibility of machine versus human value judgement, moral dilemmas and AI discrimination |
| DATA | Lack of data to validate benefits of AI solutions; Format, quantity and quality of data; Transparency and reproducibility; Insufficient size of available data pool; Lack of data integration and continuity; Lack of standards of data collection |
| TECH | Adversarial attacks; Lack of transparency and interpretability; Design of AI systems; AI safety; Specialization and expertise; Big data; Architecture issues and complexities in interpreting unstructured data; integrating legacy and new systems |
| ORG | Realism of AI; Better understanding of needs of the organizational system; Organizational resistance to data sharing and collaboration; Lack of inhouse and interdisciplinary AI talent; Threat of workforce layoffs and retraining; Lack of strategy for AI development; Embedded processes and norms; lack of resources; information silos |

That is, people learn or are transformed by observing and interacting with others. For example, in her research, success in using cloud computer services was dependent on factors such as 'external engagement' with people who were using the technology and workers' personal attitude towards risk and innovation [54]. Lastly, this paper offers a new dimension of 'transformational collaboration' (not just 'external engagement' outside the firm construct from previous research) across boundaries such as disciplines, departments, divisions, firms, industries, societies and countries which emphasizes that a person's attitude and collaboration with others who are using innovative technologies such as AI or those who have divergent backgrounds can help increase technology implementation success.

Combining a systematic and critical literature analysis with related risk models and theory, this research adds value to the field of AI by synthesizing and creating an integrated model for managing risk and benefits designed particularly for AI adoption and implementation. Using [45] as a theoretical lens, the updated McFarlan model [46] for the context of culture and social cognition theory's [53] contributions of personal attitude and engagement from previous risk research [46], "The Transformation Risk- Benefit Model of Artificial Intelligence" presents an innovative model that expands to 5 broader dimensions of 'resources' (RES), 'governance'(GOV), 'expertise building'(EXP), 'AI driven culture' (AI CUL) and 'transformational collaboration' (TR COL) that has the potential to change AI risks into benefits as illustrated in Figure 1.

## 4. MODEL PROCESSES

To fully implement "The Transformation Risk-Benefit Model of Artificial Intelligence", the authors analyze the processes that increase transformation of risks into benefits and propose the best fit for the AI context. To complete this assessment, a thorough literature review of risk processes was conducted. For example, risk processes for construction projects included identification, ranking, outcomes, monitoring and risk response [55]. To manage enterprise resource planning (ERP) implementation risks, scientists proposed 7 processes of context analysis, risk identification, risk analysis, risk evaluation, risk treatment, monitoring and review, and communication and consulting[56].

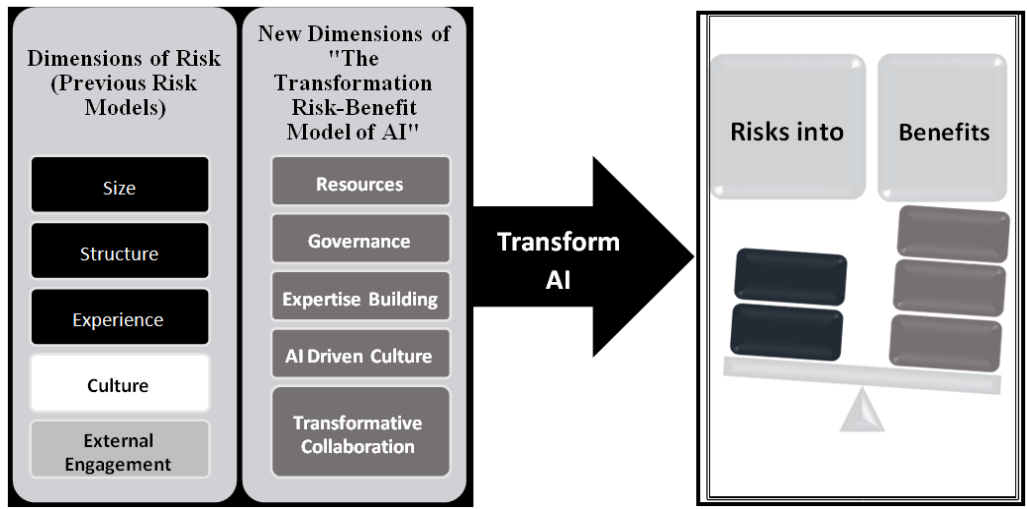

Figure 1. "The Transformation Risk-Benefit Model of AI": Transform AI risks into benefits

The railway industry uses three risk management phases: 1) risk identification, 2) risk assessment and 3) applying risk reduction plans [57]. Several actions, including risk mitigation and re-engineering to control risks, are included in traditional risk management models [58]. A four-step framework of risk assessment based on ISO standards includes (1) risk identification; (2) risk analysis (3) risk evaluation; and finally (4) risk response planning [59[60].

In medical research, there are five standard processes of conducting a risk-benefit assessment to determine the level of institutional review (IRB): 1) assess potential risks and discomforts associated with each intervention or research procedure; 2) estimate the probability that a given harm may occur and its severity; 3) explain measures that will be taken to prevent and minimize potential risks; 4) describe the benefits for subjects; and 5) discuss the potential societal benefits [61].

The general action and solution-related framework of risk management reflects that "for each issue or event requiring a decision, managers can benefit from adopting a systematic approach [60]. The main processes of risk assessment in business continuity planning are the identification of risks, their analysis and evaluation, and risk treatment and response planning [62].The drawbacks to these models are single-minded focus on risks only and not full consideration of benefits and opportunities AI can provide. After a systematic review, the researchers believe the medical model comes closest to addressing these limitations. In the same vein, an updated ISO 31000 was proposed in 2018 and expands by including how to create and protect value or benefits for stakeholders—a positive focus of AI risk-benefit assessment that was previously lacking [63].The idea is to try to identify all the consequences of a particular issue or event, to find an optimal decision set to minimize adverse effects and maximize societal and business objectives in a cost-efficient manner" [64]. This principle of 'balancing' risks and benefits is essential to "The Transformation Risk-Benefit Model of AI". Another principle of the new risk-benefit model is to be 'integrative'. Risk factors, events, benefits and consequences of AI all should be 'integrated' to meet specific organizational and societal needs. For example, in the context of the supply chain, risks could emerge from supply chain disruptions, which often create material or financial losses, delivery delays and decreased reputation and competitiveness. However, organizations that learn to successfully manage these risks can also create benefits for performance and reputation [65][66].

For small and medium sized enterprises (SMEs), bankruptcy risk is one of the most important risks identified and using a simulation model of case companies illustrates the importance of looking at 'integrative' factors. Namely, performance on several financial measures such as customer accounts receivable flow time and credit limit allowed by financial partners were predictors of bankruptcy risk which lowered the risk from 80% to 30% [67].Creating "financial sustainability" by assessing financial and nonfinancial indicators and metrics is the answer to present-day challenges of a very turbulent environment. The main goal of financial sustainability analysis is to search, discover and mobilize resources, and uncover opportunities enabling development and success of the organization [68].This approach highlights the importance of a 'strategic' outlook – what is good for the firm and society over the long haul and is another principle needed to transform risks into benefits in AI.

Suggestions made by the assessors for the improvement of the benefit-risk assessment of medical research/procedures included 4 processes: 1) A well-structured presentation of the relevant data on benefits and risks; 2) A common unit to compare benefits and risks; 3) A thinking tool/model that could add transparency; 4) A tool to help deal with uncertainty and complicated data [69]. A transformational mindset needs to 'balance' delivery discipline and accountability, agility and pragmatism, continuous improvement and a sense of chronic unease. This delicately balanced mindset enables organizations to course-correct and to quickly address emerging challenges that can be turned into competitive opportunities [70].Traditional risk models are limited by their standardized, reactive and sporadic approaches, management of risks one by one, and a myopic risk-averse mindset [71].

Due to its predilection towards emphasizing negative scenarios - not to mention its limited scope - many organizations have found traditional risk management to be lacking in its ability to provide sufficient insights and a shaky foundation on which to make informed decisions [72]. Sorting out tangled risks can only be achieved by proactively understanding and managing the attributes of interconnected risks and benefits [73].To address the limitations and drawbacks of current risk models, the authors recommend working on benefits and risks simultaneously, using five processes to 'assess' all risks and opportunities, 'identify' the level and priority of risks and opportunities, 'analyze' all the various options, plans, and strategies tied to these risks and benefits, 'create optimal plans' through collaboration across inside and outside boundaries and 'evaluate and adjust plans' on a continuous improvement loop.

The authors also add to the body of knowledge and expand their Model aimed at transforming AI risks to benefits by initiating five overriding principles of being 'holistic', 'integrative', 'strategic', 'proactive' and 'balanced'. 'Holistic' means looking at events, both positive and negative, that might impact the organization and society taking the long term big picture into account. 'Integrative' refers to seeing the synergy between solutions and across departments/divisions/industries. 'Strategic' is to have an opportunity-creating rather than a risk-adverse mindset to make plans that create outcomes to optimize the varied goals of stakeholders. A 'balanced' approach is to look at both negative and positive AI events, working out solutions that try to minimize risks, increase opportunities and transform risks into benefits. Lastly, the 'proactive' principle in the AI context is to think ahead and have plans that continually look to decrease risks, anticipate handling negative events while searching for opportunities in AI. For an overview of the principles, dimensions and processes to balance risks and benefits in "The Transformation Risk-Benefit Model of Artificial Intelligence", please see Figure 2.

## 5. PRACTICAL SOLUTIONS

Corporate commitment should extend beyond corporate interests, to address broader societal responsibilities and ethical AI practices [41]. Leaders remain in need of conceptual, technical and institutional mechanisms to assess how to achieve accountability for the harmful consequences of data-driven algorithmic systems [74]. In this section, the authors give examples of pragmatic and innovative solutions that can transform risks into benefits in each of the Model's 5 risk dimensions. Under the dimension of 'resources', innovative and practical solutions such as developing policies to assure AI will be directed at 'humanness' and common good and building inclusive, decentralized intelligent digital networks 'imbued with empathy' will help leaders meet social/ethical goals [75].

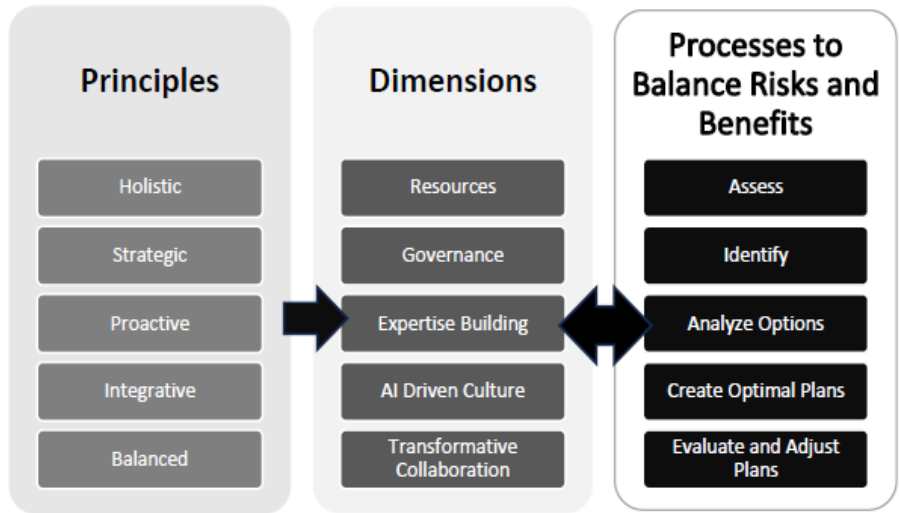

Figure 2. Principles, dimensions and processes to balance risks and benefits in "TheTransformation Risk-Benefit Model of Artificial Intelligence"

Another resource-related solution is to alter organizational, economic and political systems to better help humans 'race with the robots' and reorganize these systems toward the goal of expanding humans' capacities and capabilities [75]. Beyond retraining, a range of resource policies can help, including unemployment insurance, public assistance in finding work and portable benefits that follow workers between jobs [5]. Increasing human/AI collaboration and human interventions in decision-making in risky situations like finance, healthcare and hiring practices will ensure trust/transparency and decrease harms to people [41]. Under the 'governance' dimension, comprehensive AI lifecycle governance where policies and procedures are described and enforced throughout the AI system lifecycle can increase transparency— AI factsheets are one practical to capture metadata automatically, making enterprise validation and external regulation easier to monitor [76]. To increase fair and unbiased 'governance', create 'an auditing imperative' for algorithmic hiring systems that mandates regular internal and external audits of automated systems, based upon precedents of Occupational Safety and Health Administration (OSHA) audits in labor law or the Sarbanes-Oxley Act audits in securities law to increase trust and confidence in AI- automated systems [30].

Other effective 'governance' solutions are for organizations to conduct their own bias, explainability and robustness assessments to protect privacy and harmful outputs and security assessments to help identify and decrease threats such as cybercrimes. Also prioritize the responsible use of AI and generative AI by using "zero or first party data, keeping data fresh and well labeled, ensuring there's a human in the loop, testing and re-testing and keeping feedback close" [41]. Base 'governance' onprotecting society and the firm by following stricter regulations, guidelines & certification standardssuch as those recommended by the European Union [24].

To increase 'expertise building', train or retrain workers in new fields such as digital forensics, cybercrime specialists, data analysts and AI governance specialists, use Protective Optimization Technologies (POTs) to both explore the societal effects that algorithms and optimization systems and design countermeasures to eliminate their negative effects [77]. Widen educational opportunities to step into new AI jobs and learn new skills (throughout the job and educational systems available or create new ones), build talent in technical competencies like AI engineering and enterprise architecture and empower people to work effectively with AI-infused processes [75]. In the arena of creating an 'AI driven culture', use AI-based visualization tools [13][23], leverage open-source and user-provided data, be transparent when AI has created content using watermarks or in-app messages, cite the sources for creating content, explain why the AI gave the response it did, highlight uncertainty and creating guardrails, extend the commitment to using AI tools and technologies beyond immediate corporate interests, help meet broader societal responsibilities and ethical AI practices and prevent some sensitive, risk heavy tasks from being fully automated [41].

These practical and innovative solutions for developing an 'AI driven culture' will have important benefits such as extracting significant insights from big data and allowing workers to understand the impact of AI directly, increasing transparency, accuracy and honesty about data uses and sources, mitigating risks and biased outcomes while protecting sensitive data processes and uses [41]. In the final dimension of risk, 'transformational collaboration' solution include: 1) collaborating to minimize the size of AI models [41], 2) working with partners to create better data and train on models with large amounts of high-quality customer relationship management data; or 3) increasing collaboration, risk prevention and control across departments, disciplines, companies, countries and societies [75].

Working collaboratively can produce outstanding outcomes while lowering threats such as reducing the carbon footprint because less computation is required in smaller models, which means less energy (water and electricity) consumption from data centers and carbon emissions, higher quality data helps to maximize data accuracy, reliability and value, and working across boundaries and perspectives can facilitate the innovation of widely accepted approaches aimed at tackling wicked problems and improves risk control over complex human-digital networks worldwide [75]. Table 3 provides a sampling of practical and creative solutions organized into the five dimensions of risk.

# 6. USE CASES FOR AI

In addition to elaborating on how risks can be transformed into benefits through a sampling of doable innovative solutions, this research focuses on practical use cases in AI from three different sectors: healthcare, climate change/environment and cyber security.

## 6.1. Healthcare: Challenges and Solutions

Use cases in the healthcare sector encompass both depth and breadth in their purview including virtual care delivery, digital health driven by wearables and online medical devices, AI automation of administrative tasks, enhanced point-of-care decision systems and therapeutic access through real world data. The top 3 challenges for AI in healthcare are 1) trust; 2) selecting the right use cases and 3) managing data volume along with privacy issues [78]. Through analyzing such big data implementation cases, scientists will be able to understand how AI technical capabilities transform organizational practices, thereby generating potential benefits in industries such as healthcare [79].

One of the most recent use cases of AI's impact on healthcare is its use to fight diseases such as the COVID virus. Some generative, deep learning and machine learning methods have been used to help fight COVID-19. The main advantage of AI-based platforms is to accelerate the process of detection, diagnosis and treatment of epidemics. These AI platforms work with various data types, such as clinical data and medical imaging which improve AI approaches to find the best responses [80].

Challenges still exist, including limited data accessibility, the need for external evaluation of AI models, the lack of awareness of AI experts of the regulatory landscape of AI tools in healthcare, theneed for clinicians and other experts to work with AI experts in a multidisciplinary context and the need to address public concerns over data collection, privacy, and protection. Having a dedicated team with expertise in medical data collection, privacy, access and sharing, AI scientists handing over training algorithms to the healthcare institutions to train models locally, and taking full advantage of biomedical data can alleviate some of problems posed by these challenges and will mean AI can help create practical and useful solutions for combating pandemics [81]. Another example where AI technology is evolving in the medical/healthcare sector is using AI big data for the diagnosis, treatment and prognosis of numerous neurological diseases. Huge amounts of data are gathered from various digital sources, including electronic health records, media and databases, then analyzed, and AI algorithms are trained to detect risk factors, diagnose diseases and suggest best therapies [82]. It is estimated that AI applications can cut annual healthcare costs in the U.S. by $150 billion in 2026 by changing the healthcare model from a 'reactive' to a 'proactive' approach, focusing on health management rather than disease treatment, resulting in fewer hospitalizations, less doctor visits and less treatments [83].

AI has the potential to inspire systematic ways to process clinical and molecular information that span the 5Vs challenges of value, volume, velocity, variety and veracity, while identifying patterns of care, analyzing unstructured data, supporting decisions, prediction, and traceability of healthcare data [84].The main trend in the healthcare industry is a shift in data type from structure-based to semi-structured based such as home monitoring, telehealth, sensor-based wireless devices and unstructured data including transcribed notes, images and videos. The increasing use of sensors and remote monitors is a key factor supporting home healthcare services but the challenge is to manage the growing data generated to grow significantly and improve the quality of healthcare [85]. The challenge posed by clinical data processing involves the quantity and volume of data but also the varied types of data. AI has the potential in healthcare organizations to support clinical decision- making, disease surveillance and public health management [86]. Poor medical data processing systems are the key reasons for medical errors but using AI data management systems reduces errors and associated harms increasing patient safety and efficient data management [87]. Analyzing findings from previous clinical studies could translate new research conclusions into routine clinical processes and thus drive successful evidence-based medicine [88]. For example, Optum Labs working with Mayo Clinic, combined diverse collaborator perspectives with rich data, including deep patient and provider information, to reveal new insights about diseases, treatments and patients'behavior. Practitioners' involvement in agenda setting and continuous feedback loops improved AI-driven processes and dimensions [89]. The Rizzoli Orthopedic Institute in Bologna, Italy analyzed patients' genomic data and case histories to determine hereditary diseases risks and to provide information of effective hereditary disease treatments and develop more evidence-based surgery protocols for genetic disease patients, reducing imaging requests by 60% [90]. Being able to input clinical information into visual dashboards, increase real-time information monitoring with alerts and notifications, and real time data navigation helps healthcare leaders to make sound operational decisions [91]. With real time dashboards to monitor patients' health and prevent medical accidents, two case hospitals—Mental Health Center of Denver and Kaiser Permanente Northern California—were able to improve quality and addressed medical errors, various patient safety issues and appropriate medication use [92]. Adapting to AI-related challenges and providing constructive solutions will require a multidisciplinary approach, innovative data annotation methods and the development of more rigorous AI techniques and models. By ensuring cooperation between AI and healthcare providers and merging best practices for ethical inclusivity, software development, implementation science, and human-computer interaction, the AI community can create an 'integrated' and 'holistic' best practice framework [93]. Additionally, collaboration between multiple health care settings to share data, ensure its quality, and verify analyzed outcomes are critical to success [94].

Table 3. Practical and innovative sample solutions by risk dimensions transforming
AI risks intobenefits

| Dimensions | Practical and Innovative Sample Solutions | Transforming AI Risks into Benefits |
|---|---|---|
| RES | Prioritize people by reorganizing organizational, economic and political systems towards the goal of expanding humans' capacities and capabilities<br>Increase human/AI collaboration and human interventions in high risk situations. | Helps humans 'race with the robots' and staunches AI trends that would compromise human relevance<br>Improves trust and removes harms in finance, healthcare or hiring. |
| GOV | Build comprehensive AI lifecycle governance where policies and procedures are described and enforced during the design, development, deployment, and monitoring of an AI system.<br>Create AI factsheets to capture AI metadata across the model lifecycle automatically,<br>Create "an auditing imperative" mandate for both internal and external audits of automated hiring systems<br>Keep and audit records of job applications.<br>Use AI tools for security and threatassessments<br>Conduct bias, explainability and robustness assessments<br>Base governance on protecting society from harm and unwanted negative consequences for the firm by following stricter regulations, guidelines & certification standards recommended by the European Union<br>Test and re-test and get feedback close tosource | Builds trust in AI through transparency<br>Facilitates enterprise validation or external regulation<br>Builds trust and confidence in decision making and has been used in other areas of law, such as Occupational Safety and Health Administration (OSHA) audits in labor law or the Sarbanes-Oxley Act audit requirements in securities law<br>Helps organizations identify vulnerabilities that may be exploited by bad actors such as cybercrimes<br>Protects privacy and harmful outputs<br>Self-regulation decreases risk of harms<br>Ensure data is high quality, accurate and has value |
| EXP | Train or retrain workers in new fields such as digital forensics, cybercrime specialists, data analysts and AI governance specialists<br>Protective Optimization Technologies (POTs) systematizes the use of technologies as tools to both explore the effects that algorithms and optimization systems have on our society, and the design of countermeasures to contest their negative effects<br>Widen educational opportunities to step into new AI jobs and learn new skills<br>Building talent in technical competencies like AI engineering and enterprise architecture and training people across the organization to work effectively with AI-infused processes. | New jobs, tools and skills can help detect deep fakes, harmful patterns inthe digital arena and make workers more productive<br>Brings structure and strategies to AI tool and skill use that help not harm society<br>More educational options along with funding (grants, scholarships, internships) will increase the availability of skills throughout the organization andsociety<br>Benefits the organization by increasing efficiency and gives workers higher skilled higher paid jobs |
| AI CUL | Use AI-based visualization tools<br>Leverage open-source and user-provided data.<br>When autonomously delivering outputs, be transparent that an AI has created the content using watermarks on the content or through in-app messaging<br>Responsibly use AI in maintaining accuracy, safety, honesty, empowerment and sustainability by citing the sources from where the model is creating content, explaining why | Extracts significant value and key management information from big data<br>Ensures honesty about the use and sources of data<br>Increases transparency<br>Mitigates risks and eliminates biased outcomes.<br>Improves the accuracy of data by being transparent and in the organization and broadly across societies. |
| | the AI gave the response it did, highlighting uncertainty and creating guardrails<br>Extend commitment beyond immediate corporate interests, encompassing broader societal responsibilities and ethical AI practices<br>Prevent some tasks from being fully automated | Human intervention may protect certain risky & sensitive data processes and uses. |
| TRAN COL | Work collaboratively to minimize the size of AI models<br>Collaborate with vendors and partners to create better data and train on models with substantial amounts of high-quality customer relationship management or other generated data.<br>Collaborate across departments, disciplines, companies, countries and societies to solve bigcomplicated or wicked problems. | Reduces carbon footprint because less computation is required, which means less energy consumption from data centers and carbon emissions<br>Maximizes data accuracy, reliability and value<br>Facilitates the innovation of widely accepted approaches aimed at tackling wicked problems and maintaining control over complex human-digital networks worldwide |

## 6.2. Climate change/Environment: Challenges and Solutions

Leveraging the opportunities offered by AI for global climate change while also limiting its risks is a path that requires responsive, evidence-based and effective governance to become a competitive strategy. AI tools are transforming data-driven science — better ethical standards and more robust data curation are needed [95]. Advances in AI are helping to make sense of all the copious amounts of data and through training, machine-learning methods get better at finding patterns without being explicitly programmed to do so [96]. In environmental sciences, technologies ranging from sensors to satellites are providing detailed views of the planet, its life and its history, at all scales with AI applications for weather forecasting [97] and climate modelling [98] and for assessing damage during disasters to speed up aid responses and reconstruction efforts [95].

'Collaboration' is essential for impacting climate change and the environment since it reduces the chance of failure such as working on a problem that is not actually impactful, overly simplifying a complicated issue or using advanced computational tools when simple tools are adequate [99]. An exemplary use case is an open-source project that was sponsored and developed by the Pentagon's Defense Innovation Unit and Carnegie Mellon University's Software Engineering Institute in 2019 called xView2, which uses collaboration with many research partners. Combining machine-learning algorithms with satellite imagery identifies building and infrastructure damage in a disaster area and categorizes severity levels much faster than with current methods [100].

Specifically, this technology was used in California and Australia for wildfire responses [101], during recovery efforts after flooding in Nepal [102], and in Turkey during several earthquakes to successfully help workers be "able to find areas that were damaged that they were unaware of," in collaboration with Turkey's Disaster and Emergency Management, the World Bank, the International Federation of the Red Cross and the United Nations [103]. Although the AI technology is quite amazing, it too has problems such as its dependency on satellite images which are only clear during the day with no cloud cover and when a satellite is overhead, in remote areas there are far fewer satellite images and it does not work on the side of a building since an aerial perspective is needed. Some personnel may not trust the AI technology. New imaging called synthetic aperture radar, which creates images using microwave pulses rather than light waves, could improve on satellite technology and training of personnel to use and believe in the technology [100].

Enhancing energy efficiency can significantly contribute to reducing the impacts of climate change/environment [99]. AI applications such as smart manufacturing can reduce energy consumption, waste and carbon emissions by 30–50% and energy consumption in buildings by 30– 50%. Nearly 70% of the global natural gas industry applies AI technologies to increase weather forecasting accuracy and reliability. Combining smart grids with AI optimizes power systems' efficiency, to reduce electricity bills by 10–20%. Intelligent transportation systems can reduce carbon dioxide emissions by approximately 60%. Management of natural resources and designing resilient cities using AI can further promote sustainability [104].

## 6.3. Cyber security: Challenges and Solutions

Cybercrime is one of the most prominent domains where AI has begun demonstrating valuable inputs and is being deployed as the first line of defense. Because AI can detect new assaults faster than humans, it provides better protection against cybercrime [105]. Cyber security threat modeling with appropriate Intelligent User Interface (IUI) design helps developers enhance flexibility, usability and interaction to improving computer-human communications can reduce critical design time and create a better risk-benefit model [106]. Cyber security systems protect assets, devices, data, programs and networks from digital attacks which prevent unauthorized access [107].

Another security concept of threat modeling attempts to enhance cyber security by analyzing, itemizing and prioritizing the potential threats using the attacker's point of view [108] prioritizes strategies against an attack. These steps mirror the Model's processes of 'assessing', 'identifying', 'analyzing options' and 'creating optimal plans'. Neglecting the security designers in cyber security and focusing only on human factors in cyber-attacks [109] allows wide penetration to sensitive systems through social engineering and cyber-attacks that militarizes information technology [110]. Cyber security cuts across technical and social matters and needs to be prioritized in the movement of businesses and government activities to online platforms have changed [107]. Therefore, the use of Application Program Interface without the system engineer's control undermines the system properties and is an excellent example of the importance of being 'strategic' and 'integrative' [111].

Leaders also need to be aware of the creation of different laws and regulations with various provisions for cyber security data protection and privacy that are leading the way around the world. First is The General Data Protection Regulation (GDPR), a groundbreaking data privacy regulation created by the European Union in 2018, which revolutionized the way personal data is handled, processed and safeguarded, not only within the EU but globally. The California Consumer Privacy Act (CCPA) is another landmark privacy legislation enacted in 2020, which improves on non- discrimination, individual rights and transparency principles being used not only in California but also across the U.S. Another law includes The Cybersecurity Law of the People's Republic of China, a comprehensive legislation enacted in 2017 which safeguards cyberspace, protects individual/organizational rights and strengthens cybersecurity [112].

Predicative analytics can further improve the integration and intelligence of security design. AI-driven User Interface design must be user-centered instead of technology-centered [106]. The complexity and the number of cyberattacks is increasing daily which poses risks to both public and private technological assets [113] and must be 'proactive' and 'balance' risks with benefits [114]. Documenting findings provides a record of the risks-benefits, which can be useful for future reference or if questions arise later and to ensure that risks/benefits and strategies are not forgotten or overlooked and sharing the findings with all relevant stakeholders, including developers, managers and users ensures awareness and buy-in to achieve practical and effective solutions [114]. This is the essence of the 'transformative collaboration' dimension. Examples of AI-enabled user interface tools are Jarvis, Amazon Alexa, Netflix, IBM Watson, Nest Thermostat, Spotify and iRobot Roomba [115]. User experience, designing of user interaction based on human perception and minimizing the complexity of software integration will improve the existing network infrastructure [116] and illustrate the principles of being 'holistic', ''integrative' while enhancing the dimensions of 'expertise building' and creating a truly 'AI-driven culture'. By checking training data sets and testing algorithms for bias, leaders can 'proactively' minimize AI drawbacks [117]. "Organizations should not wait to implement AI technologies because they can be strong tools in the fight against cybercrime" [118]. Drawing from various approaches from different fields—Artificial Intelligence, Software Engineering and Computer-Human Interaction— is 'integrative' and can increase the overall usability of devices [119].

Networks are persistently exposed to threats like malware, phishing, password breaches and denial of service attacks. F-Secure Labs receives sample data of 500,000 files from its customers that include 10,000 malware variants and 60,000 malicious URLs for analysis and protection daily. They rely on AI and machine learning tools to detect suspiciousness based on the structure of a file or its behavior or both [120]. In their survey, the top-performing risk mature companies implemented twice as many of the key risk capabilities and generated three times the level of performance as those in the bottom 20%. Companies that succeed in turning risks into results create competitive 'strategic' advantage through efficient deployment of scarce resources, better decision-making, reduced exposure to negative events and applying a broad 'risk-benefit lens' to the business [50].

# 7. CONCLUSIONS

This research has systematically reviewed the most salient benefits and risks of AI and generative AI in the current literature and categorized them by their impact on society. From this analysis, examples of these challenges and benefits are categorized into 8 arenas based on recent relevant research. Next, related risk models and theories, based upon risk models in IT project management, cloud computing implementation and enterprise resource management and social cognitive theory applied from psychology are synthesized into the context of AI resulting in the newly proposed framework called "The Transformation Risk-Benefit Model of Artificial Intelligence".

This Model presents a comprehensive framework tailored to manage the ever-increasing changes in the AI environment by integrating negative risks and positive benefits. The analysis of past research found scant models and a hyper focus on risks to date. The context of AI was not fully integrated into the previous empirically-based models. "The Transformation Risk-Benefit Model of AI" expands the constructs of risk dimensions (such as 'project size' transforming into 'resources', 'culture' changing to 'AI-driven culture' and 'external engagement' widening to 'transformational collaboration' both internally and externally to the organization) based upon a systematic review.

In addition to broadening the construct dimensions, this paper contributes new insight into both the principles and processes which will enhance the transformation of risks into competitive benefits and capacities.

To address the limitations and drawbacks of current risk models, the authors recommend working on benefits and risks simultaneously using five processes of 'assessing' AI risks and opportunities, 'identifying' the risks and opportunities, 'analyzing' diverse options, plans, and strategies tied to these risks and benefits, 'creating optimal plans' through internal and external collaboration and 'evaluating and adjusting plans' in a continuous improvement fashion. Next, the authors add to the body of knowledge by presenting samples of pragmatic and innovative solutions in each of the AI and generative AI risk dimensions that highlight how risk can transform into benefit in the latest AI technology context. Lastly, three use cases in healthcare, climate change/environment and cyber security add depth to understanding the AI context and illustrate how each sector can uniquely address the principles, dimensions and processes to transform AI risks into benefits overall.

## AUTHORS

**Professor Richard A. Fulton** (M.S., Illinois State University)has taught full time computer science and information systems courses at Troy University – e campus for the past 15 years and previously at Illinois State University. His articles have been published in The Journal of Technology Research, The Journal of Scientific Information on Political Theory, Developments in Business Simulations and Experiential Learning, and the International Journal of Innovation, Technology and Management.

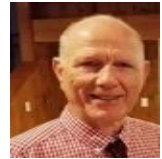

**Dr. Diane J. Fulton** (Ph.D., University of Tennessee-Knoxville) is Emeritus Professor of Management at Clayton State University, located in Morrow, Georgia. Her research interests include advanced technologies, innovations and online teaching tools. She has published several books, book chapters, and numerous articles in academic journals, including California Management Review, Planning Review, Journal of Small Business Management, International Journal of Management Education and  Entrepreneurship Theory

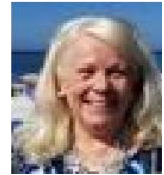

**Susan Kaplan**  (BAS, MAS, University of Minnesota-Dultuh) is Executive Vice President and Chief Management Officer of Modal Technology Corporation, a high-tech firm located in Minneapolis, Minnesota that offers new and proven solutions for artificial intelligence and machine learning that are based on modal interval arithmetic. Ms. Kaplan was Founder and President for Quality Management Solutions. She led and restructured organizations in healthcare, government, manufacturing, and service sectors to improve profitability. She is the author of The Grant Writing Process.

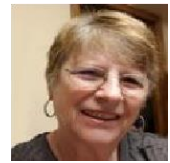

**Nathan Hayes,** CEO and Founder at Modal Technology Corporations, is an entrepreneur, mathematician, and software architect with more than 20 years of experience working in these combined fields. Hayes specializes in the applied science of modal interval analysis to the fields of artificial intelligence, machine learning and high performing computing.

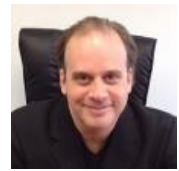